# OCCIA LAB

*To discover the causes of social, economic and technological change*

# National debts and government deficits within European Monetary Union: Statistical evidence of economic issues

Mario COCCIA

CNR -- NATIONAL RESEARCH COUNCIL OF ITALY

&

ARIZONA STATE UNIVERSITY

# National debts and government deficits within European Monetary Union: Statistical evidence of economic issues

*Mario Coccia*[1]
CNR -- National Research Council of Italy & Arizona State University

*E*-mail: mario.coccia@cnr.it

Current Address: Coccia*LAB* at CNR -- National Research Council of Italy
Collegio Carlo Alberto, Via Real Collegio, n. 30, 10024-Moncalieri (Torino), Italy

Mario Coccia  : http://orcid.org/0000-0003-1957-6731

**Abstract**.

This study analyzes public debts and deficits between European countries. The statistical evidence here seems in general to reveal that sovereign debts and government deficits of countries within European Monetary Unification- in average- are getting worse than countries outside European Monetary Unification, in particular after the introduction of Euro currency. This socioeconomic issue might be due to Maastricht Treaty, the Stability and Growth Pact, the new Fiscal Compact, strict Balanced-Budget Rules, etc. In fact, this economic policy of European Union, in phases of economic recession, may generate delay and rigidity in the application of prompt counter-cycle (or acyclical) interventions to stimulate the economy when it is in a downturn within countries. Some implications of economic policy are discussed.

**Keywords**: National Debt, Public Debt, Debt Crises, Deficit, European Monetary Unification, European Union, Economic Recession, Monetary and Fiscal Policy, Economic Growth.

**JEL codes:** E00; H60; H62; H63; H69; F43; F44; O52.

---

[1] I gratefully acknowledge financial support from the CNR - National Research Council of Italy for my visiting at Arizona State University (Grant CNR - NEH Memorandum n. 0072373-2014 and n. 0003005-2016) where this research started in 2016.

> A state debtor to foreigners is a serious evil,
> this economic fact is so evident that does not require proof.
> (Galiani F., 1780, p. 361 from the book *Della Moneta,*).

**Introduction**

Public debt encompasses all the liabilities that are debt instruments owed by governments and public administrations, public companies and other economic subjects of nations (Barro, 1979). Public debt is also a vital instrument for governments to finance public expenses, especially when it is difficult to increase taxes and/or reduce expenditure (Gnegne and Jawadi, 2013). However, a high public debt is also a critical problem for countries with weak economic system because may generate economic instability and sovereign debt crisis (Domar, 1944; Hall and Taylor, 1993; Amaral and Jacobson, 2011). In addition, high Public Debt-to-GDP ratio of countries is considered an economic issue for investors and policymakers, since it can negatively affect capital market and, in the long run, reduce investments, employment and economic growth (Coccia, 2013).

The vast literature in public economics and political economy of growth has analyzed several factors of the public debt across countries over time (Égert, 2015; Buiatti *et al.,* 2014; Elgin and Uras, 2013). However, the precise evolution of public debt between countries within and outside European Monetary Union and related European policies, before and after the introduction of the Euro currency, is overlooked. In light of the continuing importance of economic analyses concerning the evolution of public debt in current economies, this study seeks to clarify patterns of public debt across European countries to shed some empirical light on recent trends. This study focuses specifically on the following research questions:

- How is the evolution of public debt across European countries, before and after the introduction of the Euro currency?

- Have countries within European Monetary Union an evolution of the public debt similar or different to other countries?

This article endeavors to explain whenever possible, these research questions with statistical analyses. Results can clarify the evolution of public debt of European countries for supporting an appropriate political economy of growth directed to support economic growth[2] and stability of European economic system as a whole over the long run.

**Theoretical framework**

High public debt and large fiscal deficit are a common feature among countries in Europe (Tamegawa, 2016). Nations with high public debt can have problems in international lending if they do not support a sustainable commitment to repay the lenders in order to keep interest rates low on sovereign debt. The credibility of sovereign debt depends not only on the reputational consequences of borrowers but also on institutions that might prevent default from occurring (Coccia, 2017i). However, the solvency and liquidity of nations cannot solve problems of high sovereign debt, because creditors in international financial markets do not have the means to seize the assets of a defaulting borrower (Elgina and Uras, 2013; Melina et al., 2016).

Lane (2012) claims that European countries have different debt histories. Gogas et al. (2014, p. 1) argue that several European nations in the last decades have had sovereign debt crises and have faced the threat of default, such as Greece. In particular, the financial crisis from 2008 have affected the Eurozone and, combined with economic recessions, increased the public debt ratio from 67.4% in 2008 to 93.9% of Gross Domestic Product (GDP) in the 2014. Greece has reported larger-than-expected increases in fiscal deficits and elevated from 107.9 to 174.1% public debt ratio over 2008-2014 period (Lane, 2012). Ireland has increased this ratio from 27.5 to 123.7%. Italy, Belgium, Spain, France, etc. have also increased levels of public debts-to-GDP ratios in the same period (Matesanz and Ortega, 2015; cf. also, Buiatti et al., 2014; Alesina, 1988 for Italian case study). Other countries, such as Germany and Austria have experienced softer deterioration on their public debt positions, whereas Norway has experienced a reduction in its public debt stock (Matesanz and Ortega, 2015). Neaime (2015, p. 2) argues that the accumulated EU's national debts are the result of both economic and political/institutional factors. Baxter (1871) is one of the first scholars that analyzed the evolution of public debts across European countries and the pressure of public debt upon the population. Baxter (1871, p. 48) argued, about countries of the Southern Europe, that:

> the history of the debts of the Latin nations…their people are careful and frugal, but their rulers are too often reckless and spendthrift, prone to overspend their income in time of peace and still more largely in time of war; … and sometimes unable to pay even the interest. Perhaps their tendency to arbitrary and therefore irresponsible government has too much to do with the succession of deficits

---

[2] About the role of new technology, R&D investment and research labs for economic growth, see: Aghion and Howitt (1998), Calabrese et al., 2005; Calcaltelli et al., 2003; Cavallo et al., 2014, 2014a, 2015; Chagpar and Coccia, 2012; Coccia, 2001, 2002, 2003, 2004, 2004a, 2005, 2005a, 2005b, 2005c, 2005d, 2005e, 2005f, 2005g, 2005h, 2006, 2006a, 2008, 2008a, 2008b, 2009, 2009a, 2009b, 2009c, 2009d, 2010, 2010a, 2010b, 2010c, 2010d, 2010e, 2010f, 2011, 2012, 2012a, 2012b, 2012c, 2012d, 2013, 2013a, 2014, 2014a, 2014b, 2014c, 2014d, 2014e, 2014f, 2015, 2015a, 2015b, 2015c, 2016, 2016a, 2016b, 2017, 2017a, 2017b, 2017c, 2017d, 2017e, 2017f, 2017g, 2017h, 2017i, 2017l, 2018, 2018a, 2018b, 2018c, 2018d, 2018e, 2018f; 2018g; Coccia and Bellitto, 2018; Coccia and Bozeman, 2016; Coccia and Cadario, 2014; Coccia and Finardi, 2012; Coccia et al., 2010, 2012, 2015; Coccia and Rolfo, 2002, 2007, 2008, 2009, 2010, 2013; Coccia and Wang, 2015, 2016; Rae (1834); Benati and Coccia, 2017.

Current public debt histories of some countries seem to be similar to those of about 150 years ago described by Baxter (1871). As a matter of fact, the recent high debts and financial crises of some countries have generated damages on European and world economy due to weak and unstable public sectors' finances. Sargent (2012) claims that the high sovereign debt can contribute to maintain persistently high unemployment in Europe (*cf.,* Coccia, 2013). Policy makers and economists have thus been recently devoting efforts in trying to predict financial and debt crises before they occur, given the potential damage on several economic systems. In particular, in the presence of debt crises in Europe, the solvency of some European countries has become a major source of concern for the European Union (EU) as a whole, which is endangering its financial/economic integration efforts, and the successful monetary unification. In addition, some scholars suggest that austerity measures of countries may not resolve the problem of high public debts and should be accompanied with other political/institutional corrective measures (Coccia, 2017i, 2013). Matesanz and Ortega (2015, p. 757) construct a network of public debt-to-GDP quarterly ratios from 2000 to 2014 and show, in times of crisis, that:

countries' public debts tend to synchronize their changes, increasing global synchronization and hence dramatically decreasing the number of communities in the network.... as a result . . . a homogenization in the member's co-movements, producing in this way a network topological organization highly susceptible to spread the effects of the crisis among the countries. Finally, at the onset of the financial crisis the new network arrangement that appears seems to be directly related to the debt-to-GDP level itself which clearly puts into difficulties for controlling the public debt dynamic.

Many studies have focused on determinants of sovereign debt defaults, the implied interest rates paid on sovereign debt and impact of high public debt on patterns of economic growth (Elgin and Uras, 2013; Barro, 1974; Dell'Erba et al., 2013; Modigliani, 1961). Other studies have investigated possible non-linear relations between public debt and growth as well as discussed to what extent debt accumulation has a detrimental and causal effect on GDP growth (see, Panizza and Presbitero, 2014). Reinhart and Rogoff (2010) pointed out that public debt as a share of GDP may have a detrimental effect on the rate of growth of real GDP; in particular, public debt-to-GDP ratio higher than 90%, can slow down economic growth considerably (cf. also Coccia, 2013). Eberhardt and Presbitero (2015) find some support for a negative relationship between public debt and long-run growth of countries (cf., Égert, 2015). In addition, endogenous growth models suggest that public debt has generally a negative effect on long-run growth (Barro, 1990). In particular, high public debt can limit the effectiveness of productive public expenditures on long-run growth (Teles and Mussolini, 2014), create uncertainty or expectations of future financial repression (Cochrane, 2011), and increase sovereign yield spreads (Codogno et al., 2003) leading to higher real interest rates and lower private investment (Laubach, 2009).

Economic literature considers different approaches to limit government deficit and public debt, based on Neoclassical, Keynesian and Ricardian School of economics. Many studies analyzed the way how government budget deficits should be financed: e.g., by increasing taxes and/or by issuing new debt (Gogas et al., 2014; Eichengreen and Panizza, 2016). Teles and Mussolini (2014, p. 1) propose a theoretical model of endogenous growth that show how the level of the public debt-to-gross domestic product (GDP) ratio can negatively impact the effect of fiscal policy on growth. This effect occurs because government indebtedness extracts a portion of young people's savings to pay interest on debts. Therefore, the payment of debt interest requires an allocation exchange system across generations. Moreover, the large amount of debt across most developed countries also raises the discussion that cutting taxes should take into account. Although a tax cut is expected to improve long-run situation, it will undoubtedly lead to a worsening of the short-run debt situation. Tsuchiya (2016) suggests that an economy with a higher population growth has more room for a tax cut while satisfying its long-run government budget constraint. The dynamic effect of a tax cut improves the government budget situation in the long run but it is likely that low population growth leads to the deterioration of a long-run government budget. Concerning the solution of issuing new debt, Stasavage (2016) argues that states had the best access to credit when institutions gave government creditors privileged access to decision making, while restricting the influence of those who paid the taxes to reimburse debts. This situation sometimes does not improve the welfare of countries and create latent social and political issues.

Several governments and institutions in Europe, in order to reduce large government deficits, expenses and high public debts of nations, support specific measures and austerity programs based on Maastricht Treaty, the Stability and Growth Pact, the new Fiscal Compact in the Economic and Monetary Union (EMU), the creation of fiscal policy committees, etc. (Gnegne and Jawadi, 2013). However, the precise effect of these measures on the evolution of public debt across different economic systems in Europe, in the presence of financial turmoil and market turbulence, is uncertain and hardly known. The main aim of this article is to analyze the evolution of public debt and fiscal deficits across countries *within* and *outside* European Monetary Unification, during the period preceding and successive the introduction of the Euro currency. Results can provide insights on recent trends of public debt to support an appropriate political economy of growth.

**Materials and methods**

*Measures*

Macroeconomic variables under study, considering the dataset by Eurostat (2016), are:

- General government gross debt as a % of the GDP
- General government deficit/surplus as a % of the GDP
- Current taxes on income, wealth, etc. as a % of the GDP
- Taxes on production and imports as a % of the GDP
- General government fixed investment as a % of the GDP
- Total unemployment rate %
- Crude rate of natural change per 1000 persons

Data are over 1995 – 2014 period (Eurostat, 2016).

*Data Analyses and Procedures*

Countries analyzed in this study are divided in two *clubs*:

- Countries *within* European Monetary Union and with E.U. political economy (CEC): Austria, Belgium, Finland, France, Germany, Greece, Ireland, Italy, the Netherlands, Portugal and Spain.
- Countries *outside* European Monetary Union (CNEC): Sweden, United Kingdom, Denmark, Norway and Poland.

*Remark*: European Union (E.U.) political economy is based on Maastricht Treaty, the Stability and Growth Pact, the new Fiscal Compact in the Economic and Monetary Union (EMU).

The statistical analysis in the sets of countries just mentioned (CEC and CNEC) is performed considering two sub-periods, before and after the introduction of the Euro currency, *i.e.*,

- *Before* - Euro Currency period (BEC): 1995-2000
- *After*- Euro Currency period (AEC): 2001-2014

Considering the theoretical framework in economic literature, the focal hypothesis **HP** of this study is:

- **HP:** Countries *within* European Monetary Union and with E.U. political economy, from 2001, have deteriorated and increased public debt and government deficit in comparison to Countries *outside* European Monetary Union.

This study endeavors to support this HP with statistical evidence.

In particular, a preliminary analysis is performed with trends and bar graphs considering the arithmetic mean of variables across countries *within* and *outside* European Monetary Union, before and after the introduction of the Euro Currency.

The main statistical analysis is performed with regression analysis, by applying the linear model as follows:

$$Y_{i,t} = \lambda_0 + \lambda_1 T + u_{i,t} \quad i=1, \ldots, n \ (countries)$$

where:

- Dependent variable $Y_{i,t}$ = general government gross debts as a % of the GDP
- $T$=time, which is the explanatory variable
- $U_{i,t}$ is error term

The goodness of fit is performed with the coefficient of determination $R^2$. The relationships are estimated with Ordinary Least Squares (OLS) method.

The statistical analysis is also performed by applying the Independent Samples *T* Test (a parametric test) that compares the arithmetic means of two independent sets (Countries *within* European Monetary Union *vs.* Countries *outside* European Monetary Union) in order to determine whether there is statistical evidence of significant difference of arithmetic means between these two *clubs* of countries.

In particular, before and after the introduction of Euro currency, the null hypothesis ($H_0$) and alternative hypothesis ($H_1$) of the independent samples *T* test are given by:

$H_0$: $\mu_1 = \mu_2$     $H_1$: $\mu_1 \neq \mu_2$

with:
$\mu_1$ = arithmetic mean of General government gross debt as a % of the GDP in Countries *within* European Monetary Union

$\mu_2$ = arithmetic mean of General government gross debt as a % of the GDP in Countries *outside* European Monetary Union

*Mutatis mutandis,* the ANOVA for General government deficit/surplus as a % of GDP.

Statistical analyses are performed by using the IBM SPSS Statistics ® 21.0

**Results of economic facts**

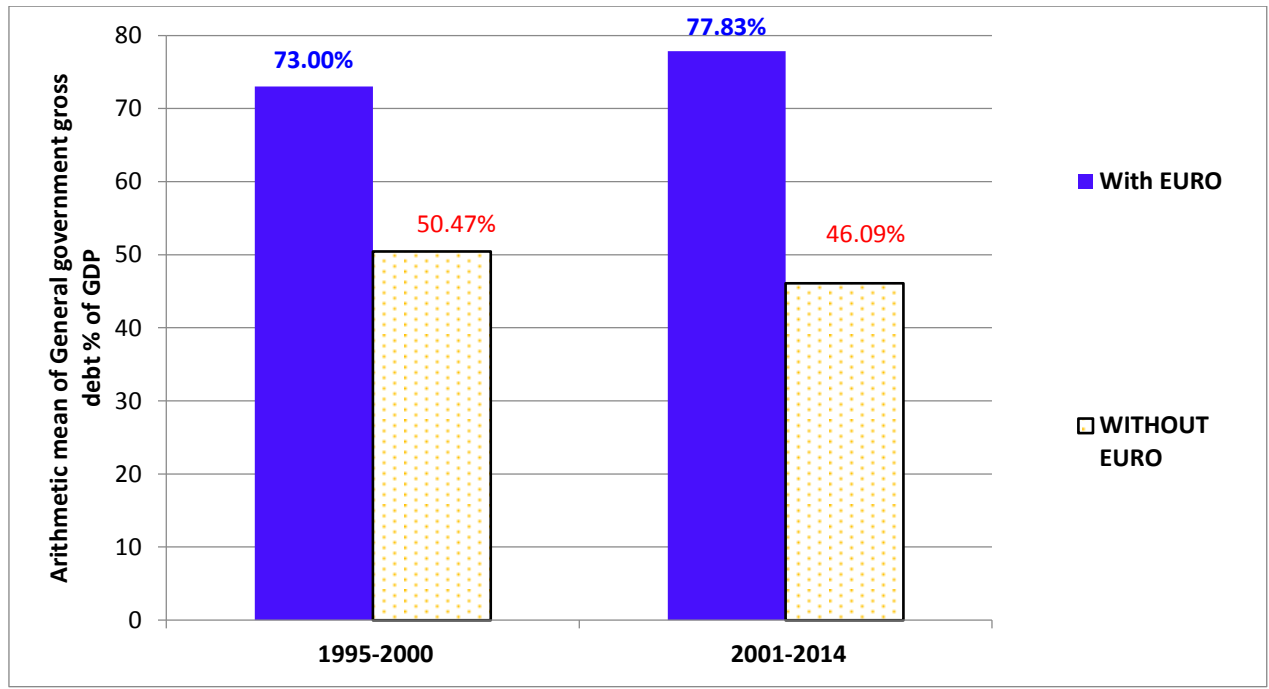

**Figure 1.** Arithmetic mean of General government gross debt as a % of the GDP of countries within and outside European Monetary Union.

*Note*: a) Countries *within* European Monetary Union: Austria, Belgium, Finland, France, Germany, Greece, Ireland, Italy, the Netherlands, Portugal and Spain. b) Countries *outside* European Monetary Union: Sweden, United Kingdom, Denmark, Norway and Poland. *Source*: EUROSTAT (2016).

Figure 1 shows that Countries *within* European Monetary Union have increased the General government gross debt as a % of the GDP from 73% in the period before the introduction of Euro Currency to about 78% in the period after the introduction of Euro Currency; whereas, Countries *outside* European Monetary Union have experienced a reduction from 50.47% to 46.09% in the same period. Figure 2 shows that Countries *within* European Monetary Union have a growing trend of General government gross debt as a % of the GDP, especially from 2007 onwards, that may be due to negative effects of the economic turmoil on economic system of these countries: in average, every year these countries have an expected increases of General government gross debt as a % of the GDP by somewhat of 1.40% ($R^2 = 0.377$, this value indicates the proportion of explained variation in total variation, *see* Table 1A in Appendix). Instead, Figure 2 shows that countries *outside* European Monetary Union have stationarity in the evolution of General government gross debt as a % of the GDP from 1995 to 2014 (regression equation is not significant statistically, see Tab. 1A in appendix).

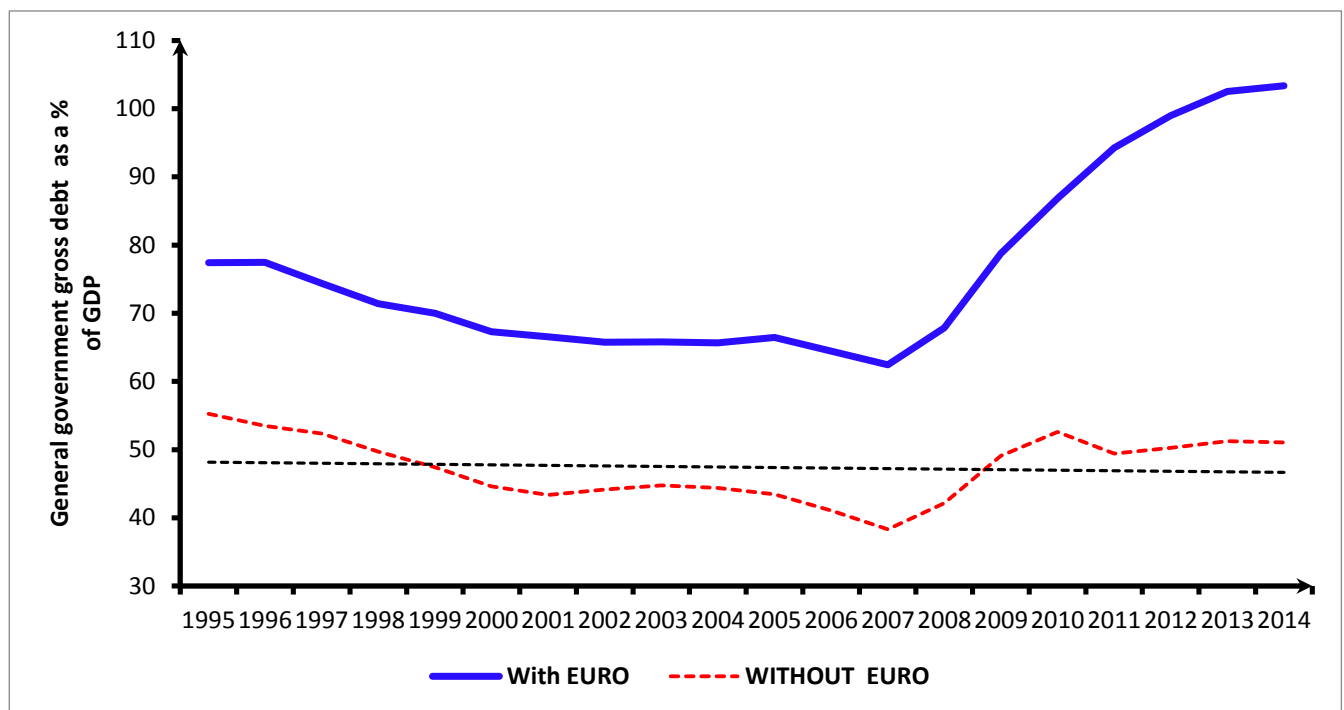

**Figure 2.** Trends of General government gross debt as a % of the GDP (1995-2014) within and outside European Monetary Union.
*Note*: a) Countries *within* European Monetary Union: Austria, Belgium, Finland, France, Germany, Greece, Ireland, Italy, the Netherlands, Portugal and Spain. b) Countries *outside* European Monetary Union: Sweden, United Kingdom, Denmark, Norway and Poland. *Source*: EUROSTAT (2016).

Figure 1A in Appendix confirms that General government gross debt as a % of the GDP of selected European countries within European Monetary Union has growing trends (e.g., Ireland, Portugal, Greece, Italy, etc.), whereas countries outside European Monetary Union have a stability of trends concerning public debt (e.g., Poland, Denmark, Sweden, etc.). Figure 3 shows that the dynamics of General government deficit/surplus as a % of GDP of Countries *within* European Monetary Union is worse than Countries *outside* European Monetary Union (2003-2014 period). In particular, Table 1 shows that average General government deficit/surplus as a % of GDP (2008-2014) is −1.77 for Countries *outside* European Monetary Union, whereas this level is considerably higher for Countries *within* European Monetary Union (−5.30). Figure 2A in Appendix confirms the heterogeneity of these trends across countries *within* and *outside* European Monetary Union.

**Table 1.** Descriptive statistics

| Variable and statistics | Countries **WITHIN** European Monetary Union | Countries ***OUTSIDE*** European Monetary Union |
|---|---|---|
| ☐ General government deficit/surplus as a % of GDP after the introduction of Euro currency (*Arithmetic mean 2008-2014*) | −5.30 (2.17) | −1.77 (2.35) |
| ☐ Current taxes on income, wealth, as a % of GDP (*Arithmetic mean 2008-2014 with base 2003=100*)* | 104.08 (4.19) | 102.83 (1.62) |
| ☐ Taxes on production and imports as a % of GDP (*Arithmetic mean 2008-2014 with base 2003=100*)* | 98.52 (3.19) | 96.06 (0.67) |
| ☐ General government fixed investment as a % of GDP (*Arithmetic mean 2008-2014*) | 3.26 (0.46) | 3.99 (0.13) |

*Note:* the different base for some indicators is for creating a comparable framework of data between the two sets of countries. a) Countries *within* European Monetary Union: Austria, Belgium, Finland, France, Germany, Greece, Ireland, Italy, the Netherlands, Portugal and Spain. b) Countries *outside* European Monetary Union: Sweden, United Kingdom, Denmark, Norway and Poland. Data before the introduction of the Euro currency are not available for these indicators in Eurostat.
Standard Deviation (SD) is in round parentheses. * The method considers a base 2003=100 for creating a comparable framework. *Source*: EUROSTAT (2016).

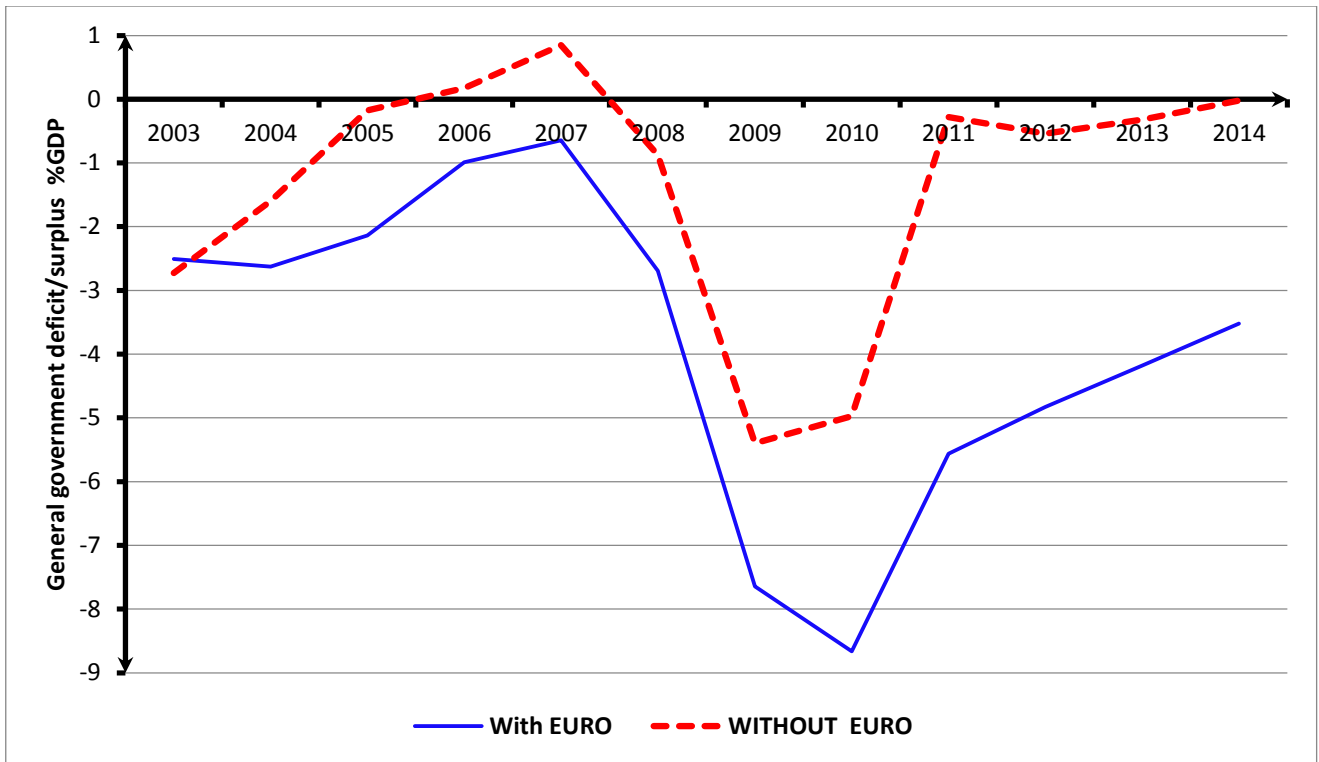

**Figure 3.** Trend of General government deficit/surplus as a % of GDP within and outside European Monetary Union, after the introduction of the Euro currency. *Note*: a) Countries *within* European Monetary Union: Austria, Belgium, Finland, France, Germany, Greece, Ireland, Italy, the Netherlands, Portugal and Spain. b) Countries *outside* European Monetary Union: Sweden, United Kingdom, Denmark, Norway and Poland. *Source*: EUROSTAT (2016).

The negative tendencies of countries within European Monetary Union over 2008-2014 may be due to manifold factors, such as high levels of current taxes on income as a % of GDP, that can affect aggregated demand during a period of economic recession. Table 1 shows that Countries *within* European Monetary Union have an arithmetic mean of current taxes on income as a % of GDP equal to 104.08, whereas countries *outside* European Monetary Union have and index of 102.83 (the method considers a base of 2003=100 to create a comparable framework).

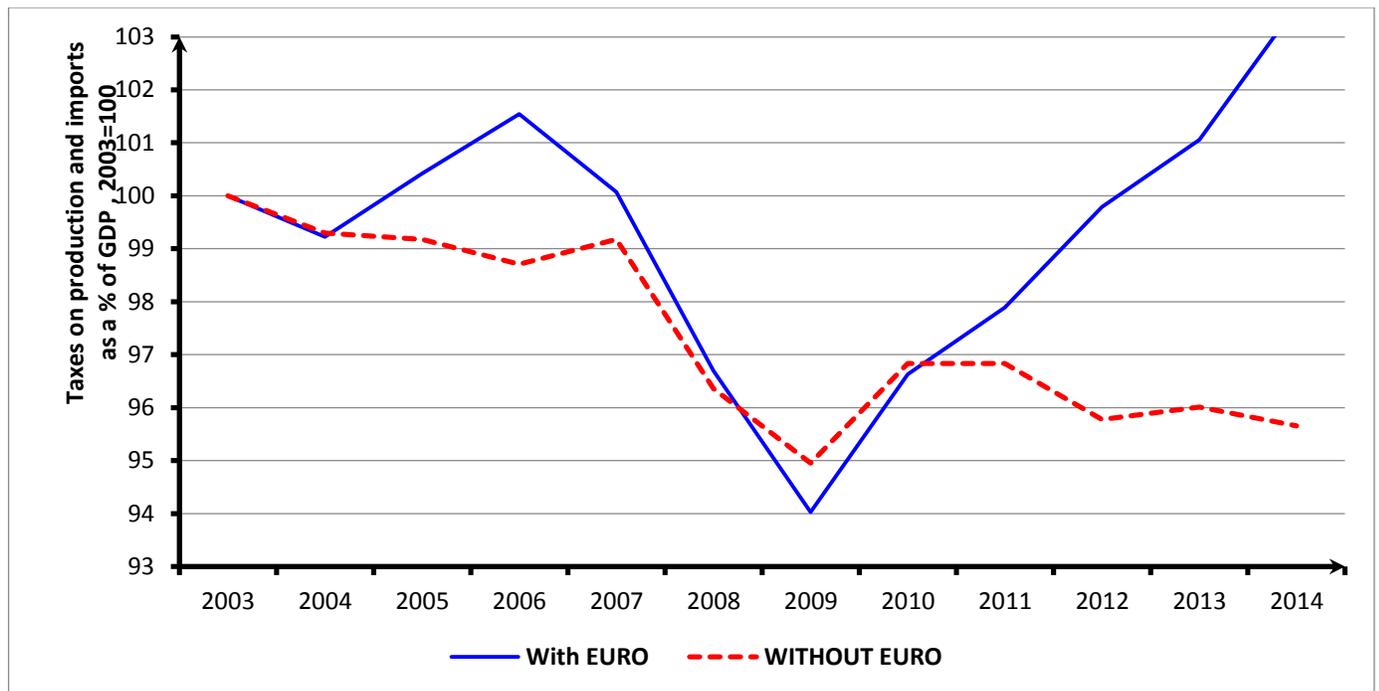

**Figure 4.** Trends of Taxes on production and imports as a % of GDP (base 2003=100) within and outside European Monetary Union, after the introduction of the Euro currency. *Note*: a) Countries *within* European Monetary Union: Austria, Belgium, Finland, France, Germany, Greece, Ireland, Italy, the Netherlands, Portugal and Spain. b) Countries *outside* European Monetary Union: Sweden, United Kingdom, Denmark, Norway and Poland. *Source*: EUROSTAT (2016).

Table 1 also shows that countries *within* European Monetary Union have a level of Taxes on production and imports as a % of GDP higher than countries *outside* European Monetary Union (98.52% *vs.* 96.06%, respectively, with 2003=100). Figure 4 shows that the trend of Taxes on production and imports as a % of GDP of countries *within* European Monetary Union has sharply increased from 2009 onwards in comparison to Countries *outside* European Monetary Union. This result may be due to negative effects of economic recessions.

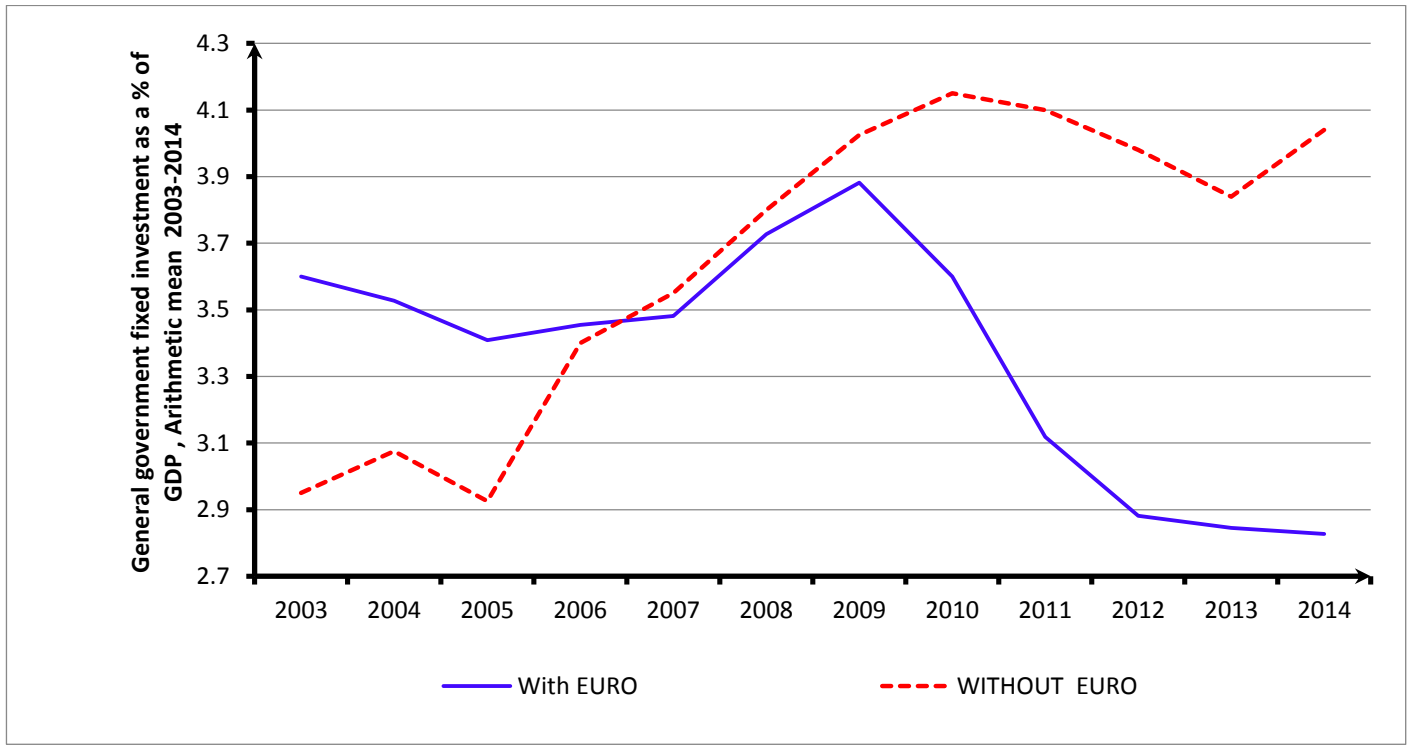

**Figure 5.** Trends General government fixed investment as a % of GDP, 2003-2014, within and outside European Monetary Union, after the introduction of the Euro currency. *Note*: a) Countries *within* European Monetary Union: Austria, Belgium, Finland, France, Germany, Greece, Ireland, Italy, the Netherlands, Portugal and Spain. b) Countries *outside* European Monetary Union: Sweden, United Kingdom, Denmark, Norway and Poland. *Source*: EUROSTAT (2016).

In addition, Table 1 shows that levels of General government fixed investment as a % of GDP in countries within European Monetary Union (+3.26%) are lower than Countries *outside* European Monetary Union +3.99% (cf., Fig. 5 and Fig. 3A in Appendix). The lower level of countries *within* European Monetary Union can be due to E.U. political economy based on Maastricht Treaty, Stability and Growth Pact, new Fiscal Compact in the Economic and Monetary Union. The different economic policies applied by countries within and outside European Monetary Union may negatively affect the evolution of public debt and its determinants. Figure 6 focuses on EU-19 countries considering trends of key variables (public debts, taxes on income, unemployment and crude rates of natural change per 1000 persons) with 2004=100 as a base for creating a comparable framework. This result may further explain possible causes of the evolution of public debt in Europe. In particular, over the 2004-2015 period, public debts are increasing in the presence of a deterioration of unemployment rate. These negative effects are combined with increases of taxes on income and reductions of growth rates of population (c.f., Tsuchiya, 2016; Coccia, 2014). In short, in the presence of these demographic, economic, fiscal and labor dynamics, a future scenario for several European countries might be a possible deterioration of structural indicators with negative effects on the European economy as a whole.

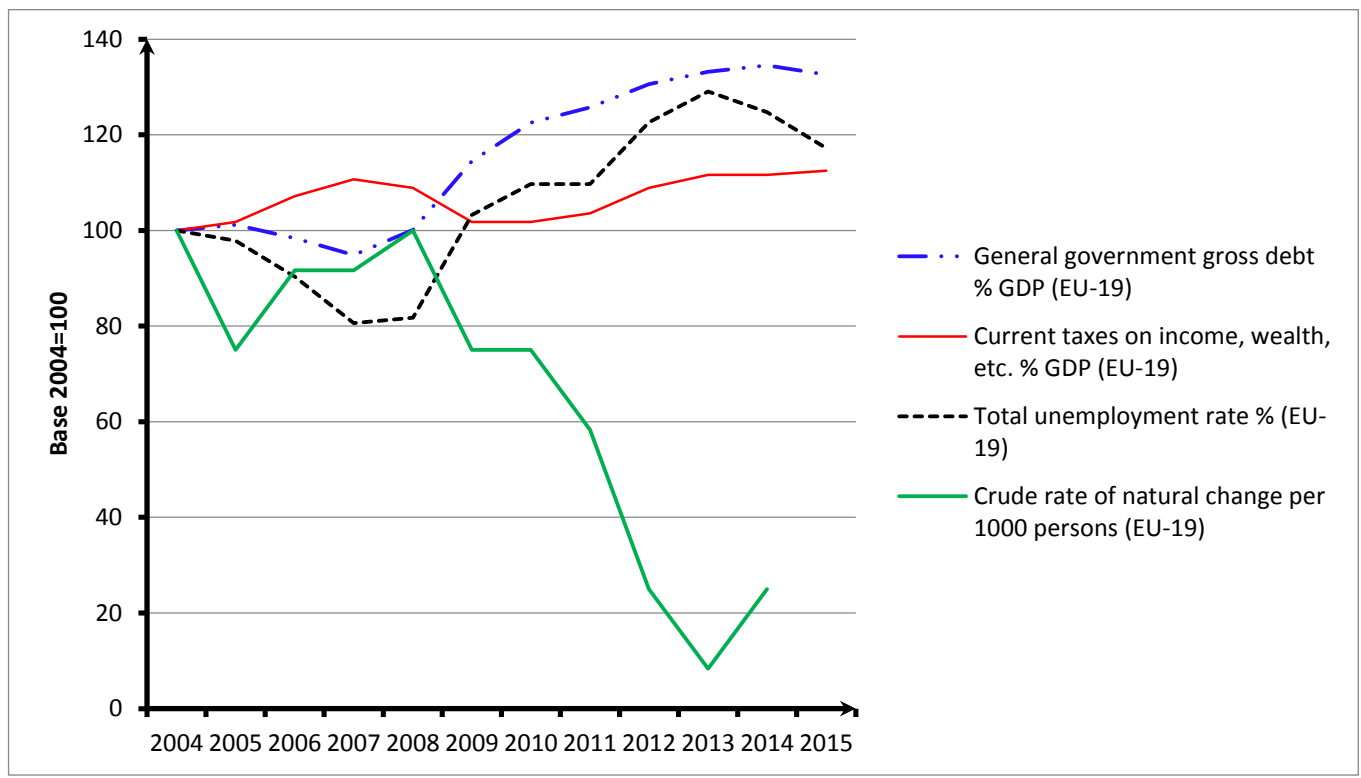

**Figure 6.** Trends of public debts, current taxes, unemployment rates and rates of natural change of population within EU-19 over 2004-2015 (2004=100). *Source*: EUROSTAT (2016).

**Table 2.** Independent Samples *T* Test with the arithmetic mean of General government gross debt as a % of the GDP (*before* the introduction of the Euro currency: 1995-2000 period).

| Countries | Years | Arithmetic mean of General government gross debt as a % of the GDP | Std. Deviation | | Levene's Test for Equality of Variances | | t-test for Equality of Means | | |
|---|---|---|---|---|---|---|---|---|---|
| *Within* European Monetary Union | 6 | 72.997 | 4.14 | | F | Sig. | t | df | Sig. |
| *Outside* European Monetary Union | 6 | 50.467 | 3.99 | Equal variances assumed | 0.04 | 0.85 | 9.59 | 10 | 0.001 |
| | | | | Equal variances not assumed | | | 9.59 | 9.99 | 0.001 |

*Note*: a) Countries *within* European Monetary Union: Austria, Belgium, Finland, France, Germany, Greece, Ireland, Italy, the Netherlands, Portugal and Spain. b) Countries *outside* European Monetary Union: Sweden, United Kingdom, Denmark, Norway and Poland. *Source*: EUROSTAT (2016).

Tables 2-3 confirm differences of General government gross debt as a % of the GDP between countries *within* and *outside* the European Monetary Union. In general, Countries *within* European Monetary Union mainly elevated its public debt ratio, whereas Countries *outside* European Monetary Union experienced a reduction in its public debt (over 1995-2014). In particular, Table 2 shows that the *T*-test for Equality of Means has $p<0.001$: *i.e.,* the statistical analysis of Independent Samples Test rejects the null hypothesis and concludes that, between countries within and outside European Monetary Union, there is a significant difference of arithmetic mean concerning General government gross debt as a % of the GDP *before* the introduction of Euro currency. Countries within European Monetary Union had an arithmetic mean of General government gross debt as a % of the GDP equal to about 73, an higher level than countries outside European Monetary Union (arithmetic mean 50.57 over 1995-2000).

**Table 3.** Independent Samples *T* Test with arithmetic mean of General government gross debt as a % of the GDP (*after* the introduction of the Euro currency: 2001-2014 period).

| Countries | Years | Arithmetic mean of General government gross debt as a % of the GDP | Std. Deviation | | Levene's Test for Equality of Variances | | t-test for Equality of Means | | |
|---|---|---|---|---|---|---|---|---|---|
| | | | | | F | Sig. | t | df | Sig. |
| *Within* European Monetary Union | 14 | 77.83 | 15.86 | | | | | | |
| *Outside* European Monetary Union | 14 | 46.09 | 4.43 | Equal variances assumed | 31.47 | 0.001 | 7.21 | 26 | 0.001 |
| | | | | Equal variances not assumed | | | 7.21 | 15.02 | 0.001 |

*Note*: a) Countries *within* European Monetary Union: Austria, Belgium, Finland, France, Germany, Greece, Ireland, Italy, the Netherlands, Portugal and Spain. b) Countries *outside* European Monetary Union: Sweden, United Kingdom, Denmark, Norway and Poland. *Source*: EUROSTAT (2016).

Table 3 shows results of General government gross debt as a % of the GDP *after* the introduction of the Euro currency. The statistical analysis of Independent Samples Test also rejects the null hypothesis and concludes of significant differences of the arithmetic mean between the two *clubs* of countries (the t-test for Equality of Means has a *p*<0.001). In particular, Countries *within* European Monetary Union have an average level of General government gross debt as a % of the GDP roughly of 78 (an higher level than period preceding the introduction of Euro currency!), whereas Countries outside European Monetary Union have an arithmetic mean of about 46.09 (2001-2014), a softer reduction of their public debt level in comparison to period preceding the introduction of Euro currency (*cf.* Tabb. 2-3).

Table 4. Independent Samples *T* Test with arithmetic mean of General government deficit/surplus as a % of GDP (2003-2014 period).

| Countries | Years | Arithmetic mean of General government deficit/surplus as a % of GDP | Std. Deviation | | Levene's Test for Equality of Variances | | t-test for Equality of Means | | |
|---|---|---|---|---|---|---|---|---|---|
| *Within* European Monetary Union | 12 | -3.83 | 2.48 | | F | Sig. | t | df | Sig. |
| *Outside* European Monetary Union | 12 | -1.32 | 2.02 | Equal variances assumed | 0.52 | 0.477 | -2.72 | 22 | 0.01 |
| | | | | Equal variances not assumed | | | -2.72 | 21.13 | 0.01 |

*Note*: a) Countries *within* European Monetary Union: Austria, Belgium, Finland, France, Germany, Greece, Ireland, Italy, the Netherlands, Portugal and Spain. b) Countries *outside* European Monetary Union: Sweden, United Kingdom, Denmark, Norway and Poland. *Source*: EUROSTAT (2016).

Table 4 analyses the average level of General government deficit/surplus as a % of GDP after the introduction of the Euro Currency. *T*-test for Equality of Means has $p<0.01$ and then this test rejects the null hypothesis and concludes that the average General government deficit/surplus as a % of GDP between Countries *within* European Monetary Union and *outside* European Monetary Union has a significant difference. In particular, Countries *within* European Monetary Union have a General government deficit/surplus as a % of GDP equal to −3.83, a higher level than Countries *outside* European Monetary Union (*i.e.,* −1.32, arithmetic mean over 2003-2014). Overall then, the statistical evidence seems in general to show the systematic differences of the evolution of public debts and fiscal deficits as a % of GDP between countries within and outside European Monetary Union: as a matter of fact, Countries *within* European Monetary Union, from 2001 onwards, have deteriorated public debt ratios and fiscal deficits in comparison to Countries *outside* European Monetary Union.

**Discussion**

Public debt of the European countries, after the economic downturn over 2008-2010 period, has sharply increased trajectory, in particular across countries within European Monetary Union (cf., Corsetti *et al.*, 2010). The high sovereign debt for some European countries contributes to maintain persistently high unemployment, low economic growth and stability of Europe as a whole (Sargent, 2012; Coccia, 2013). As a matter of fact, public debt is a complex economic issue and governments should limit public spending (Barro, 1979). Tabellini and Alesina (1990, p. 37) argue that governments choose, *a priori*, a non-optimal debt policy by budget deficits, because of disagreement between current and future majorities (such as in Italy). Corsetti *et al.* (2010, p. 45, original emphasis) claim that:

> consolidation efforts are likely to include not only tax increases but also sizeable spending cuts. . . . analysis suggests that such prospective spending cuts generally *enhance* the expansionary effect of current fiscal stimulus.

Austerity packages, balanced-budget rules and high taxation for public debt reductions in some European countries (e.g., in Greece, Italy, France, etc.) seem to negatively affect the dynamics of economic growth and public debt /GDP ratio over time, in particular in the presence of business cycle contractions and economic turmoil (cf., Afonso and Jalles, 2013; Coccia, 2013). Schmitt-Grohé and Uribe (1997, pp. 976) argue that:

> A traditional argument against a balanced-budget fiscal policy rule is that it amplifies business cycles by stimulating aggregate demand during booms via tax cuts and higher public expenditures and by reducing demand during recessions through a corresponding fiscal contraction. … an additional source of instability that may arise from this type of fiscal policy rule…, a balanced budget rule can make expectations of higher tax rates self-fulfilling if the fiscal authority relies heavily on changes in labor income taxes to eliminate short-run fiscal imbalances.

In fact, balanced-budget rule can be a source of economic instability and this result is confirmed in presence of *high* public debt that could remain constant and/or increase over time, such as in some European countries over 2005-2016 period (Schmitt-Grohé and Uribe, 1997; Stockman, 2010). Several studies do not suggest for governments a balanced-budget rule, since it affects (narrows) the political economy of driving surplus and deficits, by borrowing and lending, to smooth taxes. In addition, "the welfare consequences of decreasing ratio of debt/output at the exogenous growth rate are negligible" (Stockman, 2001, p. 439). Stockman (2004, p. 382) also claims that:

> a balanced-budget rule limits the ability to smooth taxes, rendering a large class of competitive equilibria not compatible with a government honoring its debt obligations. The reputation model predicts default as the equilibrium outcome under a balanced-budget restriction.

Although the reduction of sovereign debt of some European countries is a desirable goal for economic stability of European Monetary Unification, it is not an easy task and should be pursued by long-run economic policies, considering fluctuations of business cycles and demographic dynamics of population, to support steady-state patterns of economic growth. Europe has focused on downsizing of the ratio of public debt to GDP by higher taxation for population and balanced-budget rules for member countries. However, E.U. political economy of the Stability Pact and Fiscal Compact, in the presence of declining average rates of natural change of population, seems to generate negative effects for long-run patterns of economic growth, shaking the stability of countries and Europe to its foundations (Coccia, 2013). Antonucci and Pianta (2002, p. 306) argue that: "macroeconomic constraints of Economic and Monetary Union in Europe have put a serious limit on the economic dynamics of national economies, and of manufacturing industries in particular". De Grauwe and Fratianni (1983, p. 53) show that: "draw from the evidence . . . higher tax rates are not likely to reduce budget deficits in the long run. They may in fact increase them". As a matter of fact, results of the study here show in general that stability programmes of the European Monetary Unification seem to have deteriorated the evolution of public debts of some countries, in presence of economic turmoil and demographic crisis. Some economists show that similar rules may not be optimal policy and may generate aggregate instability, considering the initial high level of the public debt in countries (Schmitt-Grohé and Uribe, 1997; Stockman, 2010). Stockman (2004, p. 383) points out that: "The ability to borrow is desirable because debt serves as a buffer to help smooth distortionary taxes over time resulting in higher economic welfare".

Overall, then, as government consolidated gross debt is measured by the ratio government debt / GDP, an appropriate political economy of growth to reduce public debt over the long run may be to support the GDP growth (denominator of the ratio) with investments and other expansionary economic policies, rather than reduce the numerator (i.e., public debt) by higher taxation of people, also during recessions. In fact, Baxter (1871, p. 122) argued, considering England, that: "the nation by its rapid growth is constantly diminishing its burden. In the peace between 1815 and 1870 England has diminished the pressure of her debt from 9 per cent to less than 3 per cent, principally by her natural growth and increase".

**Concluding observations**

This study shows some observed economic facts concerning the evolution of public debts and fiscal deficits in Europe.

a) Firstly, differences of the evolution of public debts across European countries *within* and *outside* European Monetary Unification. In particular, General government gross debt as a % of GDP in Countries *within* European Monetary Unification has increased from 2001 to 2014 in comparison to Countries *outside* European Monetary Unification.

b) Secondly, differences of the evolution of fiscal deficits across European countries *within* and *outside* European Monetary Unification. Countries *within* European Monetary Unification experienced a severe deterioration of General government deficit as a % of GDP (arithmetic mean of −3.83 over 2003-2014) in comparison to countries outside European Monetary Unification (average value of −1.32 in the same period).

c) Thirdly, countries *within* European Monetary Unification have levels of current taxes on income and wealth as a % of GDP and taxes on production and imports as a % of GDP higher than Countries *outside* European Monetary Unification. In addition, General government fixed investment as a % of GDP in Countries *within* European Monetary Unification is lower than Countries *outside* European Monetary Unification. These effects may be due to rules of the Maastricht Treaty, the Stability and Growth Pact, the new Fiscal Compact in the Economic and Monetary Union that negatively affect the evolution of public debt and fiscal deficit across several European countries in the presence of economic turmoil and low average growth rates of population.

d) Finally, one future scenario in Europe might be that current problematic evolution of economic and demographic factors can negatively affect the performance of public debt, employment and economic growth paths for many years to come.

In general, economic theory suggests that balanced-budget rule may not be optimal policy considering the *high* initial level of public debts in several European countries (cf., Schmitt-Grohé and Uribe, 1997; Stockman, 2010). Bruno (1992, p. 204) argued: "the dangers of EMU", which are due to the decision to advance monetary union, before the fiscal federalism and coordination of fiscal, social and industrial policies across all European countries. These contradictions of the process of integration are generating negative effects on several European countries and uncertainty about future scenarios of the European Union (E.U.) economy as a whole. Perhaps the most interesting finding of this study is the high heterogeneity of structural indicators of European countries and different dynamics of public debts of European countries within and outside European Monetary Unification. However, these conclusions are of course tentative, since we know that several factors are often not equal between European countries over time and space. Results discussed here, based on aggregations of different countries, should be considered with great caution. Especially limiting is the fact that the approach here to analysis did not permit some controls and intervening variables that may have been useful in providing a deeper and richer explanation of the phenomena under study. Hence, much work remains if we are to understand in more depth the reasons of European economic policies and implications on the evolution of public debts. To conclude, more fine-grained studies will be useful in future, ones that can more easily examine other complex factors that affect the on-going public debt trends within and outside European Monetary Unification. There is need for much more detailed research into the relations between balanced-budget rule, public debt, unemployment, and economic growth.

23 | P a g e

**Appendix A.**

*Table 1A* – Regressions of General government gross debt equations

| Dependent variable | Explanatory variable: Time (1995-2014) | | | |
|---|---|---|---|---|
| | Constant $\alpha$ (St. Err.) | coefficient $\beta$ (St. Err.) | $R^2$ (St. Err. of the Estimate) | F (Sign.) |
| - Countries *within* European Monetary Unification General government gross debt as a % of GDP | −2729.44** (849.77) | 1.4** (0.42) | 0.377 (10.93) | 10.90 (0.004) |
| - Countries *outside* European Monetary Unification General government gross debt as a % of GDP | 207.04 (371.54) | -0.08 (0.185) | 0.01 (4.78) | 0.185 (0.67) |

*Note*: **Coefficient β is significant at 1%; Relationships estimated with OLS.

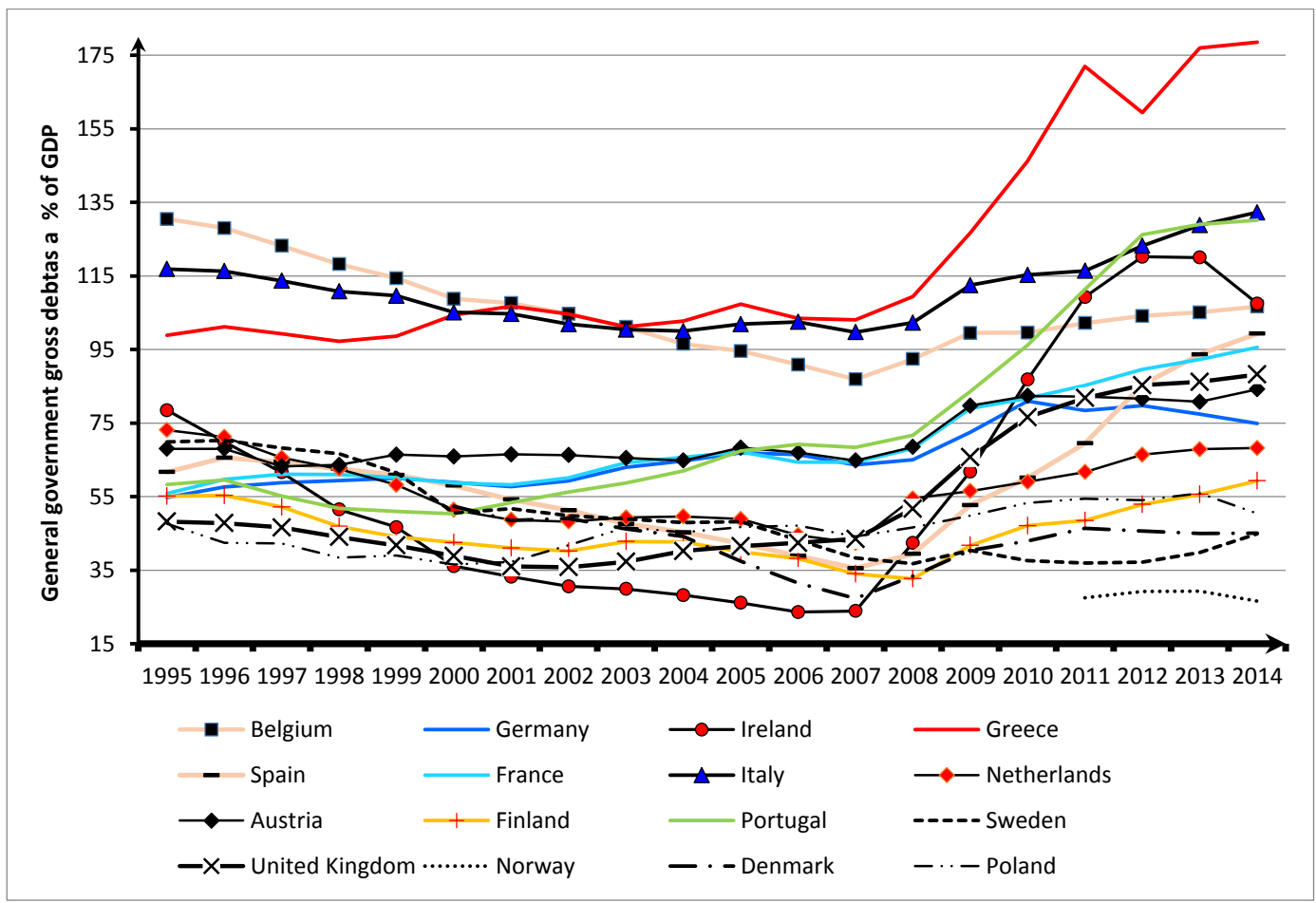

**Figure 1A.** Trends of General government gross debt as a % of GDP of selected European countries *within* and *outside* European Monetary Unification. *Source*: EUROSTAT (2016).

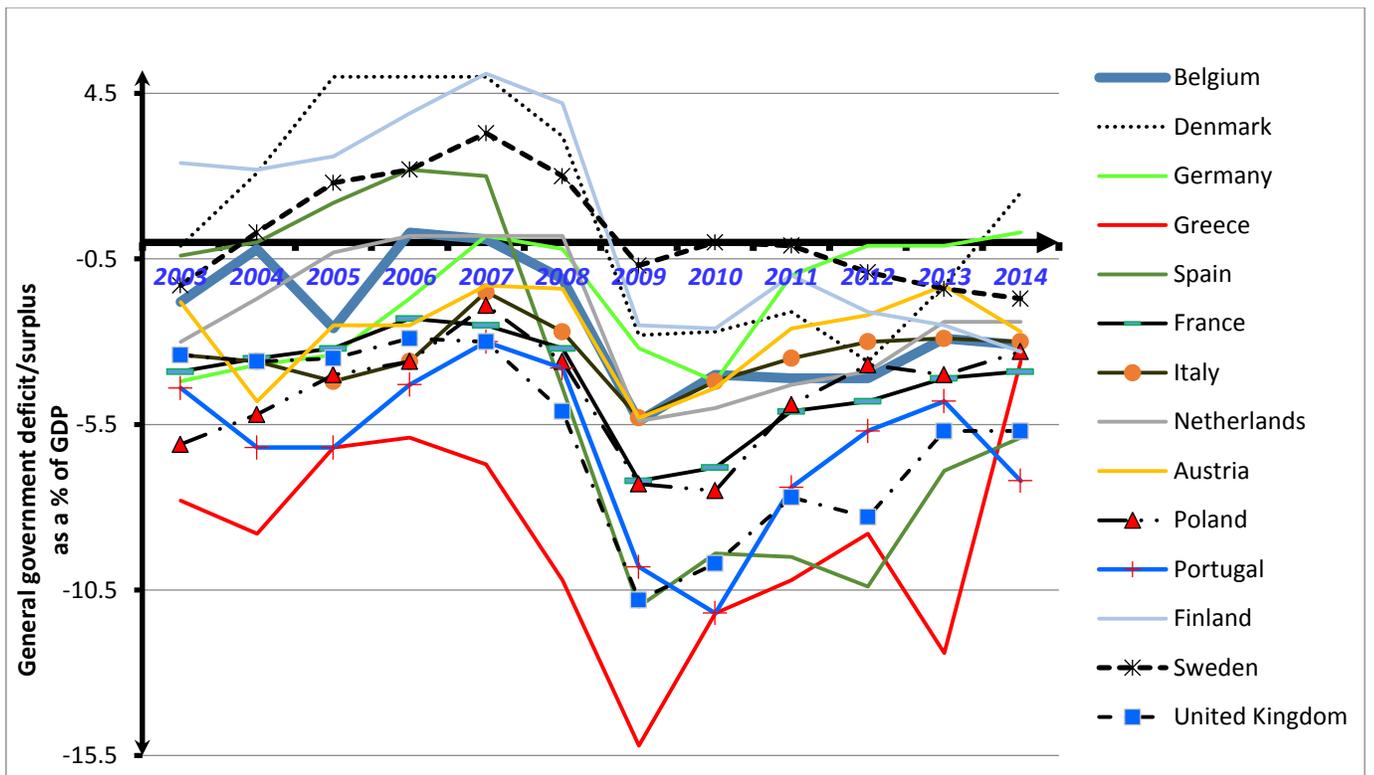

**Figure 2A.** Trends of General government deficit/surplus as a % of GDP across countries *within* and *outside* European Monetary Unification, after the introduction of Euro, 2003-2014 period. *Source*: EUROSTAT (2016).

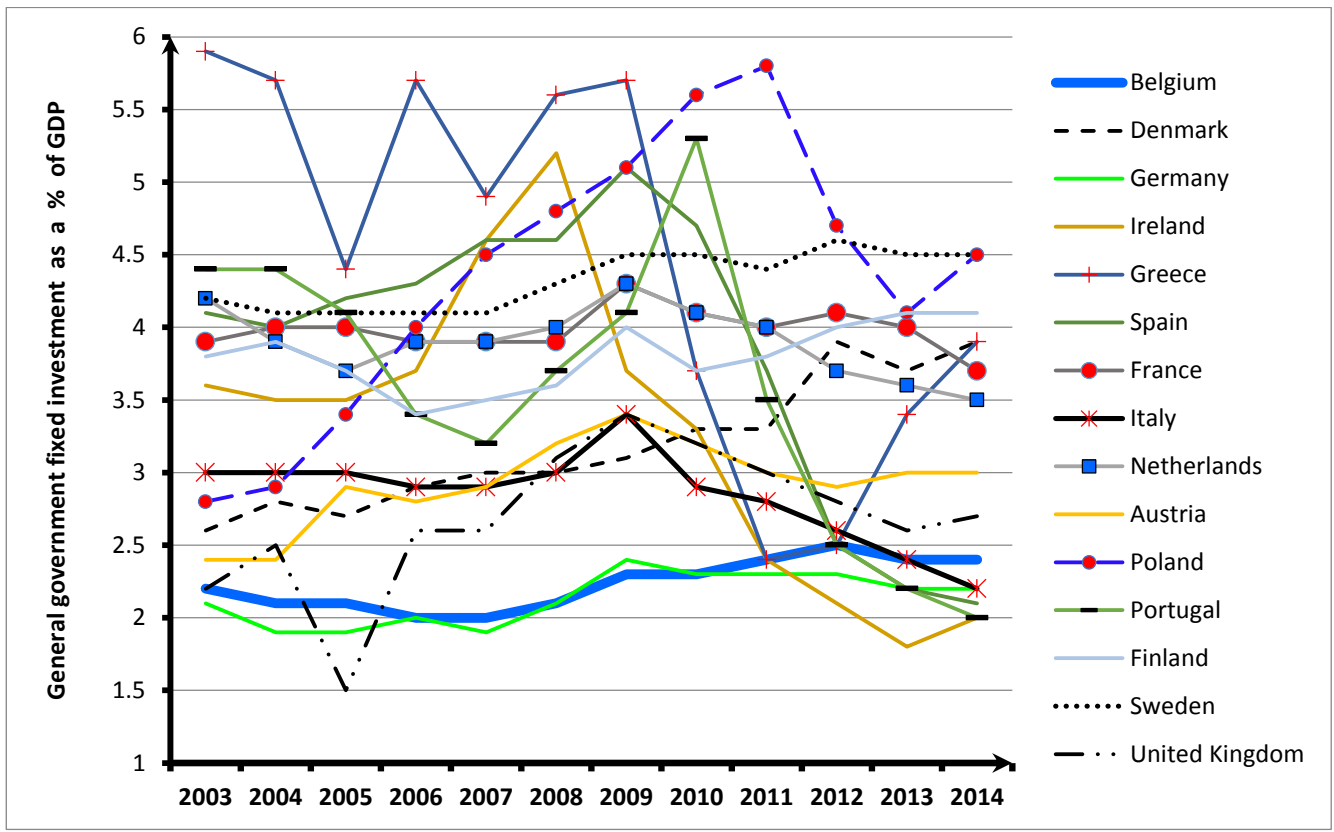

**Figure 3A.** Trends of General government fixed investment as a % of GDP between European countries. *Source*: EUROSTAT (2016).